\documentclass{prop2015}% no class options needed by now
\usepackage[english]{babel}
\def\sss{\scriptscriptstyle}
\def\^#1{^{\sss #1}}
\def\_#1{_{\sss #1}}

\def\beq{\begin{equation}}
\def\eeqno#1{\label{#1}\end{equation}}

\def\az{a_0}
\def\A{\mathcal{A}}
\def\azg{\A_0}
\def\baz{\bar a_0}
\def\l0{\ell_{0}}

\def\rar{\rightarrow}
\def\s{\sigma}
\def\l{\lambda}

\def\S{\Sigma}

\def\vr{ \textbf{r}}

\def\vv{\textbf{v}}

\def\z{\zeta}
\def\a{\alpha}

\def\c{\gamma}

\def\n{\nu}

\def\n{\nu}

\def\gN{g\_N}

\def\cmss{{\rm cm~s^{-2}}}
\def\lm{\ell\_{M}}

\def\kB{k_\text{B}}

\usepackage{epigraph}

\usepackage{etoolbox}

\makeatletter
\patchcmd{\epigraph}{\@epitext{#1}}{\itshape\@epitext{#1}}{}{}
\makeatother

\begin{document}

%%%%%%%%%%%%%%%%%%%%%%%%%%%%%%%%%%%%%%%

\title{Scale Invariance at low accelerations (aka MOND) and the dynamical anomalies in the Universe}
% \subtitle{subtitle}
\author[M. Milgrom]{Mordehai Milgrom\inst{1,}\footnote{Corresponding author\quad E-mail:~\textsf{moti.milgrom@weizmann.ac.il}}}
%\author[S.\,X. Author]{Second X. Author\inst{2}}
%\author[T.\,Y. Author]{Third Y. Author\inst{2}}
\address[1]{Department of Particle Physics and Astrophysics, Weizmann Institute, Rehovot Israel}
%\address[2]{Affiliation and address of Second X. Author and Third Y. Author}
%\shortauthors{F. Author et al.}
\begin{abstract}
Galactic systems, and the Universe at large, exhibit large dynamical anomalies: The observed matter in them falls very short of providing enough gravity to account for their dynamics. The mainstream response to this
conundrum is to invoke large quantities of `dark matter'  -- which purportedly supplies
the needed extra gravity -- and also of `dark energy', to account for further anomalies in cosmology, such as the observed, accelerated expansion.
The MOND paradigm offers a different solution: a breakdown of standard dynamics
(gravity and/or inertia) in the limit of low accelerations -- below some acceleration $\az$.
In this limit, dynamics become space-time scale invariant, and are controlled by a gravitational constant $\azg\equiv G\az$, which replaces Newton's $G$. With
the new dynamics, the various detailed manifestations of the anomalies in galaxies are predicted
with no need for dark matter.
The cosmological anomalies could, but need not have to do with small accelerations.  For example, the need for dark matter in accounting for the expansion history of the Universe is eliminated if the relevant gravitational constant is $\approx 2\pi G$. Such a `renormalization' of $G$ could be a dimensionless parameter of a MOND theory.
The constant $\az$ turns out to carry cosmological connotations, in that $2\pi \az\approx cH_0\approx c^2(\Lambda/3)^{1/2}$, where $H_0$ is the present expansion rate of the Universe, and $\Lambda$ the measured `cosmological constant'. There are MOND theories in which this `coincidence' is natural.
I draw on enlightening historical and conceptual analogies from quantum theory to limelight aspects of MOND.
I also explain how MOND may have strong connections with effects of the quantum vacuum on local dynamics.
\end{abstract}
%\shortabstract
%%%%\begin{document}
\maketitle
%%%\vskip 1truecm

\section{introduction}
\epigraph{``There can be no doubt that the interplanetary and
interstellar spaces are not empty but are occupied by a material
substance, or body, which is certainly the largest, and probably
most uniform, body of which we have any knowledge.'' }{---
\textup{J. C.Maxwell}, {\bf Ether}, Encyclopedia Britanica}

Dynamical anomalies are rife in the Universe. In all disc galaxies -- where rotation balances gravity -- the rotational speeds can be much larger than
what is dictated by Newtonian dynamics, based on the matter we see in these galaxies, made of baryons (the jargon for normal `matter'). In some such galaxies the discrepancy appears only beyond some radius, but in others it appears in full force everywhere from the center out. In dwarf-spheroidal satellite galaxies of the Milky Way and Andromeda (and probably of all galaxies), the observed characteristic speeds of constituent stars are much higher than the escape speed due to the (Newtonian) gravity of the observed mass in stars.
In all clusters of galaxies, the velocities of the member galaxies, and the temperature of the hot cluster gas --supposedly balanced by the clusters's gravity -- are much higher than Newtonian gravity would imply. And so on.
\par
In cosmology, the predictions of general relativity of the way the universe expands, and matter inhomogeneities grow, grossly disagree with observations, if baryons are the only source of gravity.
\par
The standard remedy for the anomalies in galactic systems is the invocation of large quantities of a new type of (non-baryonic) matter: `dark matter'  -- in the right amount and distribution, matched individually for each system, to account for the `missing gravity'. In cosmology, two dark components are invoked to bridge the observed discrepancies. One is `dark energy' -- whose nature remains a matter of shear speculation -- is assumed to explain the accelerated expansion of the Universe. And, cosmological dark matter is invoked to explain away other anomalies in the expansion history of the Universe, and in the formation of structure.

The MOND paradigm, in contrast, contends that there is no dark matter of a new kind in galactic systems.\footnote{A large fraction of the baryons thought to inhabit the Universe remain unaccounted for in the local Universe, known as the `missing baryons'. They could account for only a small fraction of the anomalies.} Instead, MOND posits that galactic systems are governed by dynamics other than Newtonian or general relativity.
And, a purist view of MOND, which I endorse here and elsewhere, attributes the cosmological anomalies also to an appropriate extension of galactic MOND.

Much unlike what the dark-matter paradigm does -- or can do, or will ever be able to do -- MOND predicts, and had predicted a priori, many and varied observations of galactic systems, {\it based only on the distribution of the visible baryons}.
This is because its salient predictions -- unlike those of the dark-matter paradigm -- are oblivious to the details of the formation and evolution histories of galactic systems. Like Kepler's laws for planetary systems, these predictions follow unavoidably from the modified dynamics.
\par
{\it This is the most crucial point of the comparison between the MOND and the dark-matter paradigms, and is worth enlarging on: A present day galaxy is the end product of a complicated, haphazard, and, many times, cataclysmic chain of events. These include mergers and cannibalism of other galaxies, close encounters, mass stripping, ejection of baryons by so called `feedback' due to supernova explosions or activity of black holes at the center, accretion of gas from the intergalactic medium, etc. Normal matter and the putative dark matter respond in very different ways to such occurrences. Thus, in the dark-matter paradigm, the present-day amounts and distributions of baryons and dark matter in a galaxy, are expected to be only very loosely correlated. They would also strongly depend on the exact history of the specific galaxy, which can never be known (with hardly any exception). This is why the dark-matter paradigm will never be able to predict the `dark-matter-dominated' dynamics of a given galaxy from only its baryon distribution, which is just the forte of MOND. Furthermore, the tight correlations in galaxy properties, which MOND predicts and which are found in the data, are against the expectations in the dark-matter paradigm.}
\par
There are observations that MOND does not yet account for. In its present state, it is not a complete and perfect theory.
But what existing physical theory is? As is well known, General relativity, quantum theory, and the standard model of particle physics, are all lacking in important regards. For example, the `standard model' has not been made to incorporate gravity, does not account for neutrino masses, has no room for dark matter (for those who believe in it), etc. MOND should be judged by what it does do: it does a lot, and it does it very well.
Its deficiencies should be taken as challenges pointing to necessary improvements and extensions, as is the normal way of physics.

\subsection{Dark matter?}
\epigraph{``The idols imposed upon the understanding by words are of two kinds. They are either the names of things which have no existence, ... or they are the names of actual objects, but confused, badly defined, ... Fortune, the primum mobile, the planetary orbits (epicycles), ..., and the like fictions, which owe their birth to futile and false theories, are instances of the first kind. And this species of idols is removed with greater facility, because it can be exterminated by the constant refutation or the desuetude of the theories themselves.''  }{--- \textup{Francis Bacon}, Novum Organum}
If only the cosmological anomalies are considered, invoking dark matter in the context of the `standard model of cosmology' (SMoC) is considered a success by the majority: Once its properties are rightly adjusted (its density, its being `cold'), and dark energy is also invoked with an adjusted magnitude, and with best fitting some additional parameters, one can account for the observations in cosmology (e.g., Ref. \cite{spergel15}).
\par
However, in the present context, it behooves us to note the less appealing aspects of this paradigm:
(i) There is no direct evidence for dark matter, only for the gravitational anomalies that dark matter  purports to account for.
(ii) Dark matter is needed only if we adhere to standard dynamics -- Newtonian dynamics and general relativity. (iii) All forms of matter known to exist have been excluded as dark-matter candidates, making its nature a matter of speculation beyond established physics.
(iv) Dark matter alone is not enough; another fix to standard dynamics is required -- in the form of dark energy, of an even less grounded nature and origin.
(v) Many observations conflict with the leading, cold-dark-matter paradigm (e.g., Ref. \cite{disney08,kroupa12,spergel15}). There are also some indications of inconsistencies in the SMoC itself; see, e.g., Ref. \cite{riess16} and Sec. \ref{cosmo}.
(vi) Many experiments attempting to detect dark matter directly and indirectly, at all sorts of particle-mass and -type ranges, have so far come up empty.
(vii) Intriguing, unexplained `coincidences' underlie the dark-sector paradigm: a. The required amount of cosmological dark matter is of the same order as that of the baryons (the former is about $5$ times larger than the latter), even though these two quantities were putatively determined at very different cosmic epochs, and by different physical mechanisms. b. The density of dark energy and of `dark matter' in the Universe are today of the same order. c. MOND has added to these the `coincidence' summarized in eq. (\ref{coinc}) below, between the dark energy and important characteristics of galactic dynamics. Such coincidences would be much more natural in a paradigm such as MOND, than in a scheme, such as the SMoC, where the various quantities appearing in the coincidences are, arguably, unrelated.

\section{MOND}
\epigraph{``En astronomie, nous voyons les corps dont nous \'etudions les
mouvements, et nous admettons le plus souvent qu'ils ne subissent
pas l'action d'autres corps invisibles''}{---
\textup{H. Poincar\'e}, La Science et l'Hypoth\`{e}se}

The MOND paradigm \cite{milgrom83} has been amply reviewed (e.g., Refs. \cite{fm12}\cite{milgrom14c}). So, after a succinct synopsis of MOND, I shall concentrate on some specific aspects of my choice.
\subsection{MOND synopsis}
Gravitational accelerations underlying the structure and dynamics of galaxies (and cosmology) are many orders smaller than in the solar-system. For small accelerations -- below an acceleration constant, $\az\approx 1 {\AA}{\rm s^{-2}}$, MOND dictates very different dynamics than the standard. In particular, much below $\az$ MOND dynamics are space-time scale invariant, unlike Newtonian dynamics or general relativity.

MOND has worked very well in {\it predicting} many properties of galaxies of all types and some systems of galaxies, as deduced using various probes, such as motion of stars and cold gas, hydrostatics of hot gas, light bending, etc. (with practically no free parameters): It removes the (very large) anomalies in these systems without the need for dark matter.

MOND also removes much of the anomaly in galaxy clusters, but not completely. It reduces the global discrepancy of a factor of about 7 to a factor of about 2. Various explanations for the remaining anomaly have been proposed. My favorite is the presence of yet undetected baryons in clusters. Only a small fraction of the yet unaccounted-for `missing baryons' is needed for the purpose.

We do not yet have a satisfactory picture to account for the cosmological anomalies within the framework of MOND. But, a dark-energy anomaly of the correct magnitude is naturally accounted for, at no extra cost, in some MOND theories.

MOND, as it acts in small galactic systems, may have strong conceptual connections with cosmology, because it turns out that $\az$ is of the order of some cosmologically significant accelerations parameters.

We have several full-fledged effective theories, relativistic and nonrelativistic, constructed on the basic premises of MOND (reviewed recently in Ref. \cite{milgrom15a}).  But we do not yet have one that we are fully satisfied with, and that we can consider the `final' MOND theory.

The MOND paradigm is based on the following tenets: a. The dynamics governing galactic systems that evince large mass anomalies are space-time scale invariant. Such systems are said to be governed by the deep-MOND limit. b. The boundary between the deep-MOND limit and standard dynamics is defined by a new constant with the dimensions of acceleration, $\az$. The logic behind these tenets and how they lead to the rich MOND phenomenology have been described in detail, e.g., in \cite{milgrom14,milgrom15}.

\subsection{Scale invariance at low accelerations}
\begin{figure}
\includegraphics[width=1.0\columnwidth]{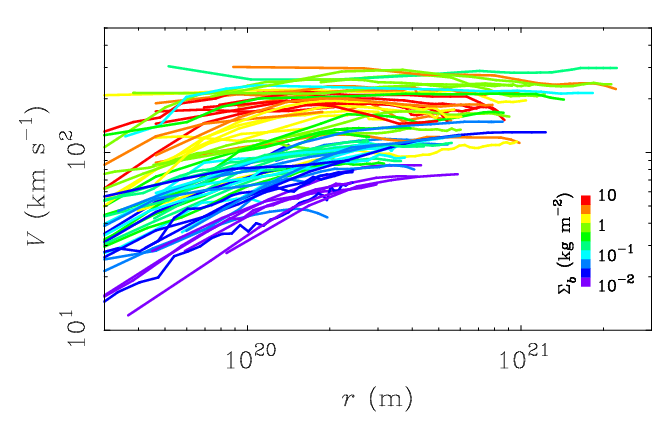}
  \caption{\label{fig1}\col
  A compilation of many rotation curves of disc galaxies, showing their becoming flat at large radii (color coded by surface density $\equiv$ mean acceleration). From Ref. \cite{fm12}.}
\end{figure}

Dynamics of gravitating systems that show large mass discrepancies (synonymous with low accelerations in MOND) are scale invariant according to the first basic tenet of MOND: The governing equations of motion are invariant under
\beq (t,\vr)\rar\l(t,\vr), \eeqno{scale}
for any $\l>0$. The inspiration for this requirement stems from the observation that the rotational speeds, $V(r)$, of test particles in circular orbits around spiral galaxies, become $r$-independent at large radii, $r$ (see, e.g., our Figs. \ref{fig1}, \ref{fig3}, Figs. 21-27 of  Ref. \cite{fm12}, and the figures in Ref. \cite{sn07}).
This asymptotic constancy of $V$ is, arguably, the most iconic manifestation of the dynamical anomalies in galaxies, because Newtonian dynamics predicts a Keplerian decline of the velocity. This departure occurs where, by definition, the anomaly is large, and is characterized by scale invariance, since two orbits of different $r$ but the same $V$ are related by the transformation of eq.(\ref{scale}).
\par
MOND extends this specific manifestation of scale invariance to arbitrary outskirt orbits, not only circular ones, and including photon trajectories in the context of light bending. And, importantly, MOND extends the principle to the many astrophysical systems that show large anomalies within their very bulk, such as low-surface-brightness disc galaxies, and dwarf spheroidal satellite galaxies of larger galaxies, such as the Milky Way and Andromeda.
\par
Gravitational dynamics that are scale invariant cannot involve Newton's $G$. Its place has to be taken by a new gravitational constant, $\azg$, whose dimensions are invariant to scaling of the length-time units (unlike those of $G$). $\azg$ can be chosen to have dimensions $[\azg]=[m]^{-1}[\ell/t]^{\c}$. ($c$ is also allowed by scale invariance but does not appear in nonrelativistic dynamics.) The boundary between the deep-MOND limit and standard dynamics has to be set by a constant with no mass dimensions. This is dictated by the universality of free fall, which we want to retain even in the deep-MOND limit (it is supported by the observations). Not wishing to introduce additional constants, we construct this `boundary constant' from $\azg$ and $G$; it then has to be $\az\equiv\azg/G$. In MOND, the second basic tenet says that $\az$ is an acceleration, which sets the dimensions of $\azg$ to $\c=4$; so $[\azg]=[m]^{-1}[v]^4$. Only $\azg$ appears in the many MOND predictions that concern deep-MOND-limit systems, while $\az$ appears as a benchmark in predictions that concern the transition between the two limiting theories.
It might seem artificial to speak of $\az$ and $\azg$ as separate constants; but this is quite instructive for efficient bookkeeping, e.g., when we formally go to the deep-MOND limit (see below), and as seen from the discussion above.
\par
The role of $\az$ in MOND is similar to that of $\hbar$ in quantum theory, or $c$ in relativity, which marks the classical/quantum boundary, and which also appears ubiquitously in quantum phenomena, as $\az$ appears ubiquitously (through $\azg$) in the deep-MOND limit. For example, $\hbar$ marks the transition in the black-body function from the classical, Rayleigh-Jeans limit (where it does not appear) to the Wien, exponential tail. Similarly, $\hbar$ marks the borderline in the dependence of the specific heat of solids on temperature, between the classical, Dulong-Petit limit (where it does not appear), and the quantum expressions (e.g., a-la Einstein, or Debye).
\par
In MOND, define, for example, the MOND radius of a mass $M$ as
\beq r\_M\equiv (MG/\az)^{1/2}. \eeqno{mondmass}
If the mass is well spread beyond its MOND radius, the system is wholly in the deep-MOND limit. While if $M$ is well contained within its $r\_M$, there is a Newtonian region within $r\_M$, and a deep-MOND-limit region far outside it.
The MOND radius is thus a transition radius, as are the Bohr radius in quantum theory -- much beyond which classical dynamics are approached -- or the Schwarzschild radius in general relativity, much beyond which Newtonian dynamics is approached.
\par
The formal limit $\az\rar 0$ takes MOND to standard dynamics, as $\hbar\rar 0$ does in the correspondence principle of quantum theory. The deep-MOND limit is gotten by taking $\az\rar\infty$, with $\azg$ kept fixed (hence $G\rar 0$).
As noted in Ref. \cite{milgrom09a}, this limit can be viewed as the limit of zero masses (as $G\rar 0$), but finite gravity, as it is now the (finite) $\azg$ that serves as gravitational coupling.

\section{Aspects of MOND phenomenology}

We have full fledged theories that embody MOND's basic tenets \cite{milgrom14c,milgrom15a}. They may differ in predicting various aspects of dynamics.\footnote{There now exist detailed N-body numerical codes, based on nonrelativistic, nonlinear MOND theories, which can be used to derive more detailed predictions of these specific theories (e.g., Refs. \cite{lueg15,candlish15}).} But they all share many robust and salient predictions that follow from only these tenets (possibly with some additional, rather plausible assumptions).
These predictions are often formulated as `MOND laws of galactic dynamics', as discussed in detail, e.g., in \cite{milgrom14}. They pertain to both general regularities and correlations for populations of galactic systems, and to the prediction of the specific dynamics of hundreds of individual systems.

Such MOND laws may be likened to the many predictions of quantum theory made before the advent of the Schrodinger equation: the black-body spectrum, the photoelectric effect, specific heat of solids, the Hydrogen spectrum, etc.
There is also an analogy to the important results (e.g., gravitational red-shift and light bending) predicted by the equivalence principle prior to the advent of general relativity.
\par
I mention here briefly just a few of these MOND laws:
 We already saw why scale invariance implies asymptotically constant rotational speeds: $V(r)\rar
V_{\infty}$ (this is the analog of Kepler's 3rd law for the deep-MOND limit).

The corresponding prediction of scale invariance for light is that the light-bending angle becomes asymptotically constant far from a mass.

\begin{figure}
\includegraphics[width=1.05\columnwidth]{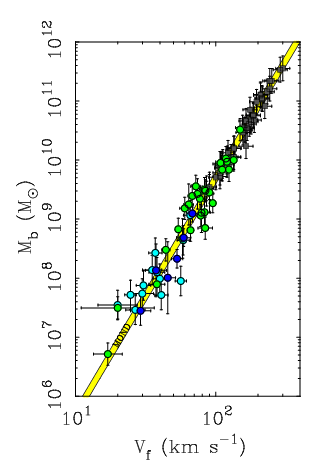}
  \caption{\label{fig2}\col
   The total baryonic mass plotted against the asymptotic rotational speed, for many disc galaxies. The (yellow) band is the MOND prediction (for a range of $\az$ values). Curtesy S.S. McGaugh.}
\end{figure}
\begin{figure}
\includegraphics[width=1.0\columnwidth]{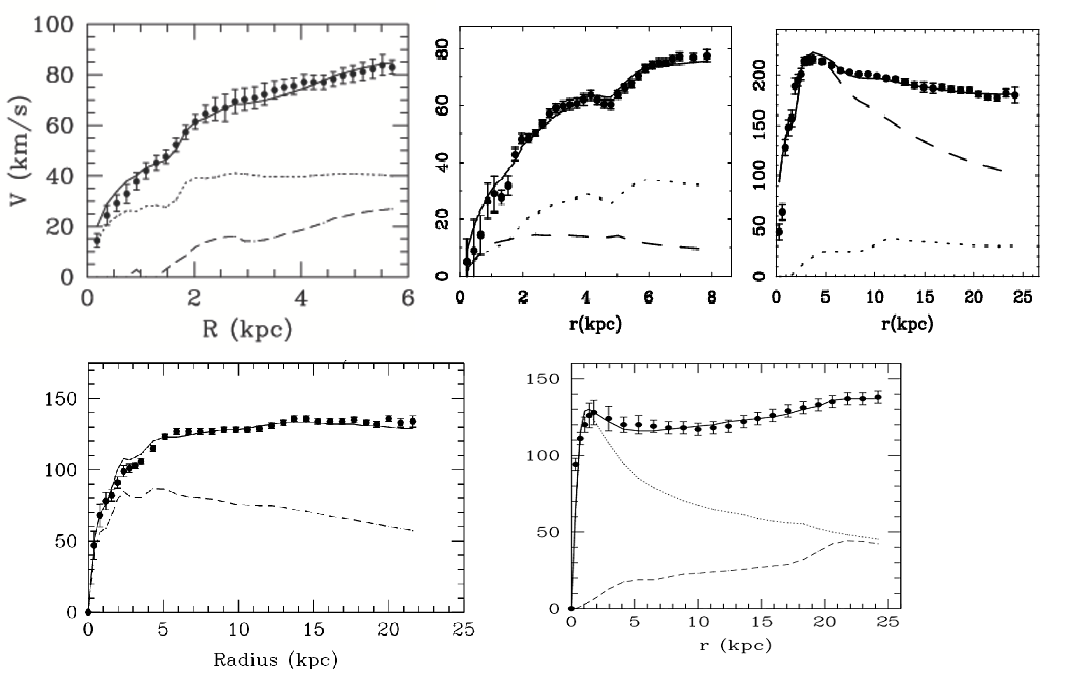}
 \caption{\label{fig3}\col The measured rotation curves (points) for five disc galaxies of varied types. The dotted and dashed lines are the Newtonian curves calculated based on the (measured) gas and/or stellar mass distributions. The solid lines are the MOND predictions for the same baryon distributions.
    Curtesy of S.S. McGaugh and R.H. Sanders.}
\end{figure}
\par
One of the most important predictions is the mass-asymptotic-speed relation (MASSR) \cite{milgrom83a} by which $V_{\infty}$ depends only on the central mass (of a galaxy, say) as:
\beq V_{\infty}^4=M\azg. \eeqno{massr}
This predicts a certain version of the `baryonic Tully-Fisher relation'. It
has been amply tested and verified for `late-type', spiral galaxies in many studies, e.g. in Refs. \cite{sanders96,mcgaugh12,papastergis16} (and see Figure \ref{fig2} here), and for a sample of `late-type' galaxies (including ellipticals) in Ref. \cite{denheijer15}. The predicted MASSR for light bending by all types of galaxies has been vindicated in Ref.\cite{milgrom13}.
\par
The predicted tight MASSR between the baryonic $M$ and $V\_{\infty}$ follows from scale invariance, while the exact power 4 follows from the role of acceleration in MOND (it is the velocity dimension in $\azg$).
In the dark-matter paradigm, $V\_{\infty}$ is wholly determined by the putative dark-matter halo. The fact that the observed MASSR is universal and applies to all galaxy types and all measurement techniques, argues strongly against dark matter. All those very different-looking galaxies have surely formed and evolved through very different histories, whose effects on baryons and dark matter were very different. Yet galaxies show this universal, tight relation between baryons and dark matter.

\begin{figure}
\includegraphics[width=1.05\columnwidth]{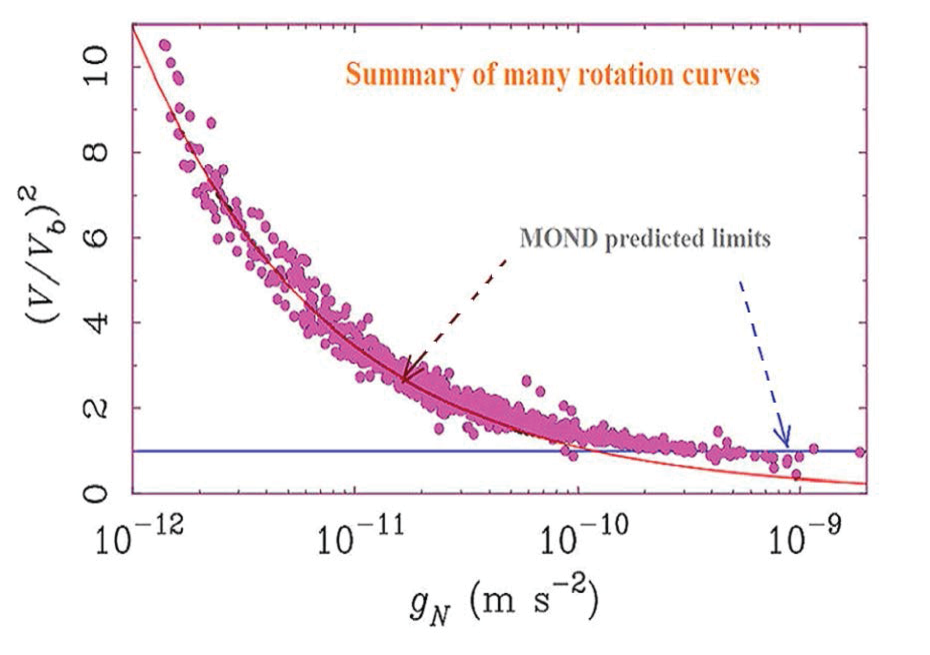}
 \caption{\label{fig4}\col
A summary of many rotation curves: The `mass discrepancy' -- the ratio, $\z\equiv (V/V_b)^2=g/\gN$ -- of the measured centripetal acceleration, $g$, to the Newtonian value, $\gN$ (with no dark matter), for many points in 73 disc galaxies (based on Fig. 10 of Ref. \cite{fm12}). The predicted deep-MOND-limit behavior (the declining line, in orange), $\z\propto \gN^{-1/2}$, is dictated by scale invariance, since under the scaling of eq. (\ref{scale}), $\z\rar \l \z$, while $\gN\rar\l^{-2}\gN$.}
\end{figure}

For the full rotation curve, $V(r)$, {\it at all radii, in all disc galaxies}, MOND predicts a strong correlation between the measured acceleration, $g(r)=V^2(r)/r$, and the Newtonian acceleration calculated from the baryonic distribution, $\gN$, (eq. (2) of Ref. \cite{milgrom83}):\footnote{This correlation becomes a functional relation for the class of `modified inertia' MOND theories discussed in Ref. \cite{milgrom94}.}
\beq  g\approx \gN\n(\gN/\az), ~~~~~\n(y\ll 1)\approx y^{-1/2}, ~~~~~\n(y\gg 1)\approx 1; \eeqno{discrep}
the $\n(y\ll 1)$ behavior is dictated by scale invariance, as $g/\gN$ scales as $\l$, under the scaling of eq.(\ref{scale}), while $\gN$ scales as $\l^{-2}$. Since $g/\gN$ is a measure of the mass discrepancy this is also called `the mass-discrepancy-acceleration correlation' and has also been amply confirmed for disc galaxies (e.g., Refs. \cite{sanders90}\cite{fm12}). A version of it is shown in Fig. \ref{fig4}. This plot encapsulates, in fact, a few sub-predictions. One is, e.g., that in all galaxies, the anomaly should set in around that radius where $V^2(r)/r=\az$.
A different, but similar-looking, correlation for spheroidal systems, as also predicted by MOND (see below), is shown in Ref. \cite{scarpa06}.
\par
Equation (\ref{discrep}) is an example of a
MOND relations that interpolates between the standard and the deep-MOND-limit regimes. It involves an `interpolation function', $\n(y)$, only the two asymptotic forms of which are dictated by the MOND tenets.\footnote{There may be other such interpolating functions in other contexts, just as there are various functions interpolating between the classical and the quantum regimes.} Its exact form has, at present, to be put in by hand, and would be a prime target for derivation in a more fundamental MOND theory. The situation is similar to that with the black-body function before Planck's quantization, when only the form for small (Rayleigh-Jeans) and large (Wien) $h\n/\kB T$ values were known.

\par
In general, scale invariance means that velocities depend only on (baryonic) masses, not on sizes [unlike standard dynamics where $V\propto(M/r)^{1/2}$].
We saw an obvious example in asymptotic flatness, where scale invariance in the deep-MOND limit is applied to different orbits in the same system.
But scale invariance has also far reaching consequences in relating different systems. For example, is implies that isolated, low-acceleration systems with the same baryonic mass have the same typical velocities, evinced, e.g., in the tightness of the MASSR (Fig. \ref{fig2}). This applies also to spheroidal systems, such as elliptical galaxies, or dwarf spheroidal satellites -- which are supported by `random' motions, not by rotation -- where MOND predicts a correlation between the characteristic  velocity dispersion $\s$, and the (baryonic) mass, $M$,
$M\azg=q\s^4$, where $q$ is of order unity, but depends on dimensionless characteristics (shape, velocity distribution, etc.) and can vary from system to system. This prediction underlies the well known observational `Faber-Jackson' relation for  elliptical galaxies, which is of this form.

As another important example: MOND predicts that for two deep-MOND-limit systems in which the observed baryon mass density distributions are related by scaling $\rho_2(\vr)=\a\l^{-3}\rho_1(\vr/\l)$ (the total masses $M_1=\a M_2$), the measured velocity distributions are related by $\vv_2(\vr)=\a^{1/4}\vv_1(\vr/\l)$.
For example, if two thin, flat disc galaxies have surface density distributions $\S_1(r)=\l^{-2}\S_2(r/\l)$ ($\a=1$), not only their asymptotic speeds are equal, but their rotation curves are predicted to be related by $V_1(r)=V_2(r/\l)$. This prediction has indeed been tested in several instances of observed pairs of galaxies. For example, in McGaugh, ``Empirical Tests of MOND in Galaxies''\footnote{$http://astro.u-strasbg.fr/MGAtotheDARK//talks/mcgaugh.pdf$}, Ref. \cite{mcgaugh14} (Fig. 3 there), and Ref. \cite{bf16} (Fig. 1 there). This is also raised as a potential difficulty for the dark-matter paradigm in Ref. \cite{oman16}, where several additional examples are shown.

Indeed, all the above phenomenology is quite against the expectations from the dark-matter paradigm. In the first place, the scalings predicted by MOND, and seen in the observations, pertain to the {\it baryonic} mass distributions, not to the putative dark matter, which supposedly dominates the dynamics for deep-MOND-limit systems. Second, the dark-matter paradigm is based on Newtonian dynamics, which is not scale invariant. So to reproduce the MOND predictions in the dark-matter paradigm would require tall conspiracy, not only in general correlations, but for each individual system, separately.
\par
Indeed, far beyond the above-mentioned, general relations and correlations, MOND predicts the full dynamics of many individual galaxies from only their baryonic mass distribution, a feat that, as mentioned, the dark-matter paradigm is inherently incapable of performing.
\par
For individual disc galaxies, MOND predicts the full rotation curves based only on the baryon distributions in them; this is the flagship of MOND predictions and testing. In some cases, undulations and other features on the observed rotation curves occur at radii where the mass discrepancy is large -- so the putative dark matter would dominate there (see some examples in Fig. \ref{fig3}). These appear in the predicted MOND rotation curves, because they result from corresponding features in the baryon distribution without `dark matter' having washed them out as a dark-matter halo would do, perforce. MOND predictions of rotation curves have been amply tested; e.g. in Refs. \cite{kent87,begeman91,dm98,sn07,gentile11,haghi16}, and many more.

For elliptical galaxies, MOND predictions can be tested, e.g., through the hydrostatics of hot gas around them. \cite{milgrom12a}. For dwarf satellite galaxies of the Milky Way \cite{serra10} and Andromeda \cite{mm13}, MOND predictions have been tested against the measured velocity dispersions of their stars.

MOND theories are generically nonlinear, even in the nonrelativistic regime. The reason is that accelerations in the deep-MOND limit have to scale as $(M\azg)^{1/2}r^{-1}$; so are not linear in the masses (in Newtonian dynamics, where $G$, not $\azg$, appears, they scale as the linear expression $MGr^{-2}$).
This nonlinearity gives rise \cite{milgrom83} to the so called `external-field effect', whereby deep-MOND-limit dynamics within a system falling in an external field are affected by the external acceleration field, even if it is constant. This has been discussed extensively; e.g., in Refs. \cite{milgrom83,mm13,wu15,hees16,haghi16}.

\section{The significance of $\az$}
The constants $\azg$ and $\az=\azg/G$ appear ubiquitously in, and are now established part and parcel of, galactic phenomenology. For example, $\azg$ is the predicted proportionality constant in the MASSR [eq. (\ref{massr}) and Fig. \ref{fig2}], and, more generally, it determines the normalization of the low-acceleration asymptote in Fig. \ref{fig4}. $\az$ appears as the intersection value of the two limits in Fig. \ref{fig4}, and both appear abundantly in rotation-curve predictions of MOND, etc. {\it Such constants do not emerge in any known version of the dark-matter paradigm.}
\par
MOND thus brings under the same umbrella disparate phenomena, all involving $\az$, which would appear unrelated without it. This, again, is analogous to the disparate quantum phenomena, all involving $\hbar$.
\par
We get from the data $\az\approx (1.2\pm 0.2)\times 10^{-8}\cmss$. As has been noted from the very advent of MOND (e.g., \cite{milgrom83}, and the detailed discussion in Ref. \cite{milgrom94}), this is a cosmologically significant acceleration. We have the near equalities
\beq \baz\equiv 2\pi \az\approx cH_0\approx c^2(\Lambda/3)^{1/2}, \eeqno{coinc}
where $H_0$ is the Hubble constant, and $\Lambda$ the observed equivalent of a cosmological constant (aka dark energy).
Thus, as mentioned above, to the list of `coincidences' that underlie the mass anomalies in the Universe, MOND has exposed another: the appearance of the cosmological acceleration parameters in local dynamics in systems very small on cosmological scales.
\par
It is also useful to think of this coincidence in terms of the `MOND length', $\lm\equiv c^2/\az$, and
the MOND mass, $M\_M\equiv c^4/\azg$. The former is of the order of today's Hubble distance:
$\lm\approx 2\pi \ell\_H$ ($\ell\_H\equiv c/H_0$), and also of the de Sitter radius, $\ell\_S$, associated with the density of dark energy (the cosmological constant) $\Lambda$: $\lm\approx 2\pi \ell\_S$. And, $M\_M\approx 2\pi c^3/GH_0\approx 2\pi c^2/G(\Lambda/3)^{1/2}$, is of the order of the closure mass within today's cosmological horizon.
\par
This numerical coincidence, in itself, has some `practical' phenomenological ramifications. For example, it implies that we cannot have local systems that are both relativistic and in the MOND regime (e.g., we cannot have a black hole that probes the MOND regime). A system of mass $M$ and size $R$, $MG/R\approx c^2$, and $MG/R^2\lesssim\az$, has $R\gtrsim\lm>\ell\_H$; so the system cannot be smaller than today's cosmological horizon.
{\it Except for the Universe at large, MOND does not have to deal with strong gravity systems}.
\par
It follows, for example, that the emission of gravitational waves by relativistic systems, such as coalescing black holes, occurs at very high accelerations ($g/\az\approx\lm/R\_G\ggg 1$, where $R\_G=2MG/c^2$ is the gravitational, or Schwarzschild, radius of the system) and would thus follow closely the dictates of general relativity.
In fact, for waves emitted by highly relativistic systems where most of the system mass participate (for example coalescing black holes of similar masses), the acceleration associated with the wave itself is $g\_W/\az\approx c^2h\_W/\l\az$, where $h\_W\approx R\_G/D$ is the strain associated with the wave, $\l\approx R\_G$ its wavelength, and $D$ the distance it traveled from its source. So, $g\_W/\az\approx\lm/D$, implying that such waves enters the MOND regime only after traveling over cosmological distances.
Other implications are discussed in Ref. \cite{milgrom14c}. Different relativistic MOND theories can make very different predictions as to the modes and propagation of gravitational waves under low-acceleration conditions (where both the environment and the wave are characterized by low accelerations). So future detailed observations of gravitational waves may help constrain such theories.
\subsection{Why an acceleration?}
Beyond such phenomenological implications,
relation (\ref{coinc}) may point to possibly the most far-reaching implication of MOND: The state of the Universe at large is somehow tightly connected with local dynamics in small systems: Either the state of the Universe affects local dynamics -- e.g., by determining inertia of bodies, in the spirit of Mach's principle -- or, the same agent enters both cosmology (in causing the accelerated expansion) and local dynamics a-la-MOND.
\par
As we live in a space-time that is not flat, but is curved,\footnote{Not to be confused with the fact that space, at any given cosmic time, is as flat as can now be measured.} with a characteristic radius $\ell\_S$,  gravity must be affected for distances of this order (as enforced, e.g., by an inclusion of a cosmological-constant term in general relativity). This is not, however, the effect in MOND, where departure from standard dynamics occur below some acceleration, not beyond a certain distance.
Nevertheless, the proximity in eq. (\ref{coinc}) may explain why it is {\it an acceleration} that marks the boundary between the two dynamical behaviors \cite{milgrom99}: In many physics contexts, an acceleration $a$ defines a physically significant scale length $\ell_a=c^2/a$. For example, this is the radius of the near field of an accelerated charge; it is the distance to the event horizon carried by a uniformly accelerated system; it is the characteristic wavelength of the Unruh radiation associated with an accelerated system \cite{unruh75} (see also Sec. \ref{fund}), etc. In a sense, then, low accelerations probe large distances.
Thus, $a\gg\az$ implies $\ell_a\ll\lm$, so $\ell_a$ does not probe cosmological distances, and is not `aware' of the curved nature of space-time. But for $a\ll\az$ it does, and may cause the dynamics to break with the standard.
\par
This situation is analogous to quantum physics of particles in a box of size $L$, which defines a borderline momentum $P_0=\hbar/L$, such that the energy spectrum for $P\gg P_0$ is practically oblivious to the presence of the box, but not so for $P\lesssim P_0$.
\par
An example of a concrete realization of this suggestion is described in Ref. \cite{milgrom99}, based on the quantum vacuum as an `absolute space' frame.

\section{\label{fund}What fundamentals underlie MOND phenomenology?}
Many questions have arisen in light of the phenomenological success of MOND, and many important ones are still moot. Here, I list some of these questions that pertain to the possible origins of MOND. Some of these were discussed in more detail, e.g., in Refs. \cite{milgrom94,milgrom99,milgrom09a,milgrom15a}.
\par
In the first place one can ask whether MOND phenomenology indeed bespeaks `new physics'; namely, the galactic anomalies are not due to the added gravity of dark matter, but to modified laws of dynamics. The alternative is that the anomalies are due to the gravity of dark matter, and MOND results somehow from the involved processes of galaxy formation and evolution, as Kepler's laws might have resulted, in principle, from the complicated history of planetary-system formation. This is a repeating claim by dark-matter advocates; but,
I and others have maintained that this is highly inconceivable (see the discussion above, and, among many, Ref. \cite{wu15}).
\par
Assuming, as I do hereafter, that new physics is involved, one needs to construct
a MOND theory, i.e., one that embodies the basic tenets of MOND described above. It is then guaranteed to produce the main phenomenological predictions of MOND. However, beyond this requirement there is much leeway in constructing theories, and this is, indeed, reflected to some extent in the number and variety of MOND schemes and full-fledged theories propounded to date.
\par
There are some very basic questions that one asks oneself when looking for a MOND theory.
For example:
Is MOND relevant only to gravitational phenomena, or is it more general, and may enter electromagnetism, for example? Recall, in analogy, that quantum theory was first thought to be relevant only for light and its interaction with matter; whereas, its relevance first to electron dynamics (Bohr atom), then to atoms (e.g., specific heat of solids), and to physics at large, was understood only years later.
\par
This question is intimately tied with the question of wether MOND is a modification of gravity or is it, more generally, `modified inertia'.
By the former we mean that only the Poisson action (in the nonrelativistic case) or the Einstein-Hilbert action (in the relativistic case) are to be modified\footnote{These actions depend only on the gravitational degrees of freedom.}. Modified inertia includes theories that modify the free actions of the various degrees of freedom (and then also, probably, of the gravitational action, since it is the `free action' of the gravitational degrees of freedom).
\par
The second major question is: ``is MOND, as we now know it, an emergent theory that follows from a more fundamental framework?'' I have argued repeatedly that this is most probably so. This is based largely on two observation: The first is relation (\ref{coinc}), which may hint that MOND somehow emerges in a larger context that involves the Universe at large.\footnote{An analogy: Galilei's acceleration constant, $g$, appearing in near-Earth-surface dynamics, the Earth's radius, $R\_{\oplus}$, and its surface escape speed, $c_e$, are related by $g=2c_e^2/R\_{\oplus}$. This relation can be understood only within the deeper universal Newtonian gravity, and indicates that constant-free-fall-$g$, Galilean theory is emergent.}
The second indication is that most existing MOND theories involve an imposed interpolation between the standard and the deep-MOND regimes.
\par
Emergent MOND would be in good company.
For example, the standard model of particle physics is believed to be an effective (or emergent) theory, in light of its above-mentioned shortcomings and the many free parameters it employs. General relativity and quantum theory are also probably approximations of a deeper-layer theory of quantum gravity. For example the Einstein-Hilbert gravitational action -- a cornerstone of general relativity, is believed to be emergent from vacuum effects -- an idea going back to Sakharov \cite{sakharov68}. It has also been suggested that Einsteinian gravity emerges from thermodynamic considerations (e.g., Refs. \cite{jacobson95,padmanabhan10,verlinde11}).
And, the possibility that quantum theory itself, apart from gravity, is emergent, has been a subject of active research since its inception.
\par
MOND could be emergent both as modified gravity or as modified inertia. For example, extending the above ideas of Refs. \cite{jacobson95,padmanabhan10,verlinde11} (or others) -- which were dealt with in a flat space-time --  to a de Sitter, curved Universe, governed by a cosmological constant, might produce MONDian gravity instead of Einsteinian one.
\par
As to modified inertia, I have argued in Ref. \cite{milgrom99} that if deep-MOND dynamics is to be understood as emergent, then, we should seek to explain the whole of inertia as emergent, with the inertia-producing mechanism having the two limits: the standard and the deep-MOND limit.
\par
One encounters in (standard) physics many instances of acquired or modified inertia (as embodied in the effective free actions). For example, the idea that inertial masses and their dependence on velocity are acquired due to electromagnetic self interactions, goes back to the turn of the 20th century (e.g., in the work of Max Abraham).
\par
Another common theme is of effective inertia that results from, or is modified by, interactions with a background medium whose degrees of freedom are integrated out. In particle physics: the
renormalization of masses, and the Heisenberg-Euler correction to the Maxwell action, are due to interactions with the quantum vacuum; the acquisition of finite masses by the intermediary bosons of the electro-weak sector via the Higgs mechanism; and the masses of the fermions due to interaction with the Higgs field. In condensed matter: the renormalization of electron masses in solids; the appearance of holes as effective particles; the appearance of emergent photon-like particles in certain quantum spin liquids; the behavior of electrons as massless, Dirac particles in graphene; and the recent emergence of effective behavior of electrons as Majorana, massless particle.\footnote{In the particle-physics examples, the effective actions gotten, which are Lorentz invariant, are extended to curved space-times through `minimal coupling': replacement, everywhere, of the Minkowski metric by the space-time metric. This ensures the universality of free fall. In the (nonrelativistic) condensed matter context, this is not relevant.}
Emergent relativistic inertia and passive gravity in the context of `acoustic space-times' have been demonstrated in Ref. \cite{milgrom06a}.
\par
It is thus conceivable that inertia in whole, as would conform with the MOND basic tenets, could emerge from
some underlying physics.
\par
As an intriguing possibility, I suggested in Ref. \cite{milgrom99} that the whole kinematics underlying inertia could somehow result from the effect of the quantum vacuum on the various degrees of freedom. The vacuum would thus serve as the proverbial, Machian `rest of the universe'. Indeed, I argued, any accelerated system `knows' that it is being accelerated with respect to the vacuum, because it is then subject to the `Unruh effect' \cite{unruh75}: a manifestation of the vacuum in systems that are not inertial with respect to it, showing as `real' radiation. For example, this radiation causes the population of internal exited states of the system.
The quantum vacuum thus provides an important prerequisite for acquired inertia -- the detectability of non-inertial motion -- and can thus potentially define `absolute space'.
\par
For a system on a not-so-realistic, eternally-constant-acceleration trajectory, the Unruh radiation is thermal, with a temperature proportional to the acceleration, $a$:\footnote{It has proven hard to calculate the effect for more general trajectories, in which case the radiation is still there, but is not quite thermal.}
$T(a) = \a\_U a$, where $\a\_U\equiv \hbar/2\pi c \kB$.
The Unruh effect refers to a system in a Minkowski, flat space-time. In the context of MOND, its generalization  to curved space-time is of prime importance, since it does exhibit, as MOND would have it, marked difference between accelerations smaller and larger than some cosmological acceleration parameter. For example, in a de Sitter Universe, which is governed by a cosmological constant $\Lambda$, and which approximates our present day Universe, this borderline acceleration for the generalized Unruh effect is
$a\_{\Lambda}\equiv c^2(\Lambda/3)^{1/2}$,
just as appears in the `coincidence' (\ref{coinc}).\footnote{See, e.g., Ref. \cite{chiou16}, for another example of how a characteristic length of the background universe can enter local measurements of the Unruh effect in accelerating systems.} Furthermore, as heuristically detailed in Ref. \cite{milgrom99}, one gets the correct two limits for accelerations below and above $a\_{\Lambda}$. The arguments are based on the fact that the Unruh temperature for a constant-acceleration system in a de Sitter background is
\beq T(a) = \a\_U(a^2+a^2\_{\Lambda})^{1/2}. \eeqno{unruh}
This idea is in the basis of several subsequent schemes to get MOND as an effective theory; e.g., Refs. \cite{pikhitsa10,klinkhamer12}.
\par
There is no contradiction between this concept of induced inertia and the idea that (some) particles get their masses through their interaction with the Higgs field, for example. The latter would then only fix the `coefficients' that determine the relative strength of inertia for the different particles or systems (and, in any event, much of this mass comes from internal interactions and kinetic energies within the compound systems, not from constituent rest masses). The fact that these `coefficients' couple to inertia, and the exact dependence of inertia on the local, and possibly global, characteristics of the motion -- such as acceleration -- would be dictated by the mechanism I discuss here. In other words, the `mass coefficients', as dictated by mechanisms such as the Higgs interactions, appear in front of the free particle actions, but the form of this action as a functional of the particle degrees of freedom would be determined by the mechanism alluded to here. This is analogous to the special-relativistic picture in which rest masses are dictated by one mechanism, and are coefficients that dictate the relative strength of inertia for different bodies, but the `kinematic' Lorentz factor appears from other considerations.
\par
There are ideas to obtain MOND phenomenology in galaxies from microphysics of a newly introduced, omnipresent medium. Two examples are the programs lead by Blanchet (e.g., Ref. \cite{blanchet15}), and by Khoury (e.g., Ref. \cite{khoury16}). In these, the medium is double purpose: It acts as dark matter on cosmological scales via the added gravity of its mass, while its microscopic properties and interactions with baryons reproduce MOND in galactic systems (where the effects of its gravity have to be suppressed).
\par
The above MOND-from-vacuum scheme -- which also hinges on the interaction of baryons with an omnipresent medium --  has, I feel, the following advantages:
a. We know, at least, that the responsible medium exists. b. The required (Unruh) effects are known to act. c. The full range of accelerations is included, with no interpolation between the Newtonian and deep-MOND-limit regimes put in by hand, albeit for limited trajectories. d. The appearance of $\az$ as having a value of order $c^2\Lambda^{1/2}$ is derived, not put in by hand. While the model remains heuristic, limited, and not yet established on concrete dynamical effects of the vacuum, it may well show the way to a full-fledged theory.
\par
The third major question that we ask is what cherished basic principles of `standard', pre-MOND physics might have to be abandoned or replaced in a MOND theory.
\par
When completing classical physics to the quantum regime (or to the relativistic regime) we had to forego, or greatly modify, some long-cherished principles: Einstein's explanation of the photelectric effect forced a view of EM radiation as quantized, despite perceived evidence to the contrary. And, QM forced us to relinquish absolute determinism in physics. Both `sacrifices' had been long resisted, as is well known.
Relativity forced us to abandon the absolute nature of time, and replaced Galilean invariance (additivity of velocities) with a new symmetry.
\par
MOND may very well entail loss or replacement of some cherished principles; but which?
We already know, for example, that the MOND external-field effect implies a breakdown of the strong equivalence principle, even in the nonrelativistic regime, as the dynamics within a small, self-gravitating system (Einstein's, proverbial, freely-falling elevator) does depend on the external acceleration.

Another potential break with accepted principles is the possibility that the Universe at large enters strongly into local dynamics. The idea itself  underlies the Mach principle, so is not quite new, but MOND gives it a new look, and strong support, in the form of relation (\ref{coinc}).

There are various principles, such as Lorentz invariance, universality of free fall, etc. that have been verified experimentally to very high accuracy, but {\it so only in the very-high-acceleration regime}. It may well be that they break down substantially in the deep-MOND limit; so we need not shy from abandoning or modifying them in MOND theories that otherwise make good sense.
For example, the deep-MOND limit may enjoy a symmetry different from Lorentz (see, e.g., Ref. \cite{milgrom06,milgrom09a}, and Sec. VII-A in ref.\cite{milgrom15a}).\footnote{As explained in Ref. \cite{milgrom94}, MOND would still require some symmetry connected with boosts, to account for gravitational light bending.}

\section{\label{cosmo}MOND and cosmology}
A purist and minimalist view would hold that the anomalies in the cosmological context should also be explained within the MOND paradigm, with no dark matter.\footnote{Unlike some suggestions, such as those described in Refs. \cite{blanchet15,khoury16}, that the two issues call for separate solutions: that some omnipresent medium acts as dark matter, by its added gravity, in cosmology, while the galactic anomalies are due to emergent MOND dynamics due to microscopics of the same medium.} This includes the anomaly associated with the accelerated expansion of the Universe, which calls for dark energy, and which, from the outset is redolent of MOND [see eq. (\ref{coinc})]: a `cosmological constant' of the magnitude observed is natural in MOND.
\par
But, there is not yet a satisfactory account, in the MOND paradigm, for the cosmological anomalies attributed to dark matter. This is often brought up as a severe challenge for MOND.
However, it has to be understood that, epistemologically, what remains to be accounted for in cosmology is much less than what MOND has already accounted for in the realm of the galaxies, with a simple set of postulates.
Each galaxy is a universe on its own in terms of the amount of independent data that need to be accounted for in connection with the dark-matter anomalies.
After all, Newtonian dynamics was established based on only one planetary system and the phenomenological laws that govern it (Kepler's).
\par
Because of the clear interconnections of MOND with cosmology, I believe that MOND cosmology will not be understood by first finding a MOND theory that works locally and then applying it to describe the Universe at large, as has been the case with general relativity. This theory, unlike MOND, is not tied to cosmology from the outset. I thus think that MOND and cosmology will have to be understood together as parts of the same theory. Nevertheless, even at the present state of MOND, we can bring up several insights that may be useful.
\par
Let us first compare the mass discrepancies and their suggested solutions (dark matter and MOND) in galactic systems and in cosmology.
In galactic systems, the mass discrepancies can be very large, as large as a factor 50 or more, and they vary greatly from system to system, and with location within a system. To account for the mass discrepancies even in just the few hundred galaxies studied to date, one needs hundreds of {\it independent} parameters to describe (fit for) their `dark-matter' halos. This is because the very turbulent and haphazard formation histories of galaxies do not permit us to determine from first principles the (dominant) effect of the putative dark matter from only its observed baryons.
MOND claims to account for all these with no free parameters.
\par
In cosmology, the anomalies related to dark matter appear observationally only during the cosmological epoch that starts somewhat before the time of decoupling of the baryons from the radiation (as a result of recombination).
\par
The cosmological anomalies whose conventional explanation requires dark matter are:
1. The total density of matter, $\rho\_M(t)$ (baryons plus dark matter), appears in the general-relativistic equations that determine the expansion history of the Universe (the time dependence of the scale of the Universe). We have a good notion of what the baryon density, $\rho_b(t)$, is. But, to get the observed expansion history, we need (in general relativity) to have
$\rho\_M(t)=\z_c\rho_b(t)$, where $\z_c$ is one of the parameters of the SMoC. The SMoC implies that $\z_c$ is a constant during the relevant period. Using the most up-to-date SMoC best fits \cite{planck15}, I find $\z_c\approx 2\pi$, within the uncertainties.
2. Explaining the exact structure of the anisotropies in the cosmic microwave background (CMB), in particular, `the height of the third peak', requires, in general relativity, the presence of neutral matter prior to the decoupling of baryons from the CMB (baryons were ionized at the time).
3. The time since decoupling is too short for the small baryon homogeneities at the time of decoupling to have lead to the present (large) inhomogeneities. Dark matter would help in that its inhomogeneities would have been more pronounced at the time of baryon decoupling.
\par
MOND can, potentially, account for the third cosmological anomaly in a natural manner (without dark matter), since the inhomogeneities are characterized by low accelerations, and would develop much faster in MOND \cite{milgrom10}.
However, we do not have the correct theory to calculate structure formation in detail. Heuristic applications of MOND did show that structures as strong as observed at present can form, but the details have not been fully reproduced yet (see, e.g.,  Refs. \cite{nusser02,llinares11,candlish16}).
MOND does not yet have a convincing answer to anomaly 2, although there are aspects of MOND that could, in principle, account for it (e.g., Ref. \cite{milgrom10}).
In what follows I concentrate on the first cosmological anomaly.
\par
Because the matter density, $\rho\_M$, appears in the equations of standard cosmology as $G\rho\_M$, the observed expansion history can also be gotten with no dark matter (i.e., $\rho\_M=\rho_b$), if the value of the gravitational constant that is relevant for the epoch after matter-radiation decoupling is not Newton's $G$, but $G_c\approx \z_c G$.\footnote{There are limits on $\z_c$:  $|\z_c-1|\lesssim 0.13$ for the much earlier epoch when the `primordial' light elements were formed \cite{carroll04}. But this period was characterized by strong domination of radiation over matter, whereas in the post-decoupling epoch rest mass of nonrelativistic matter dominated the matter energy density. So it may be that $\z_c$ had different values during the two epochs.}
\par
While in its application to cosmology, the SMoC is considered a success, potential problems have been pinpointed. One that was sharpened recently \cite{riess16} is this: The determination of the present-day Hubble constant (expansion rate of the Universe) -- a central parameter of the SMoC -- as measured from the redshift-distance relation of nearby galaxies \cite{riess16}, is at odds with the value claimed from best fit of the CMB anisotropies to the SMoC \cite{planck15}. The latter is sensitive to the whole expansion history since decoupling. So, if not due to problems with the observations themselves, such a disagreement could appear if, for example, parameters that the SMoC assumes fixed during this epoch, such as $\z_c$, or the density of dark energy, do, in fact, vary (or if other of the SMoC's building blocks need modifying).
\par
In galactic systems, MOND accounts for the anomalies as being related to low accelerations, $a<\az$, and predicts the acceleration discrepancy to be $\z\approx \az/a$.
This acceleration-discrepancy connection might apply in the cosmological context as well. Then, in looking for a MOND explanation we should pinpoint accelerations, $a_c<\az$, that characterize the discrepancy.
One should also consider the possibility that $\az$ varies with cosmic time; so for each epoch one has to compare $a_c$ with the momentary $\az$. 
\par
As a {\it heuristic} example, consider the case where $\az$ is constant, and the `cosmological' dark matter is simply the totality of `phantom matter' -- i.e., the fictitious dark matter that would mimic the effects of MOND in galactic systems, and that would appear to surround such systems. Assume that what matters are only the accelerations produced by over-density bodies, with the smooth matter distribution not entering. Let the baryon fraction
in identified galactic systems be$f\_G$, and the rest assumed in a homogeneous component.\footnote{We do not know where most of the baryons are today; so this is only an assumption.}
For a concentrated mass, $M$, the phantom halo extends to a distance from $M$ where the acceleration becomes of order of the `ambient' acceleration $g\_{\infty}$, where it is `cut off' by the external-field effect, mentioned above. The total dynamical mass of $M$ (baryons plus phantom) is then $\approx M\az/g\_{\infty}$. Then, the total density is $\rho_d\approx \rho_b f\_G \az/g\_{\infty}$, where $\rho_b$ is the baryon density. Today we have $f\_G\approx 0.1$, and $\az/g\_{\infty}\approx 50$; so we get $\rho_d/\rho_b\approx 5$, which is in the right ballpark of the required $\rho_d/\rho_b\approx 2\pi$, deduced with general relativity. However, it is not clear that this idea can be justified by some good theory of MOND that explains cosmology. For example, a. it is not clear whether the baryons that are not in structures do not contribute phantom matter of their own; b. it is not clear offhand how to estimate this ratio in all the epochs since recombination, and how to account for the fact that this ratio has been at least roughly constant during this time; c. MOND also predicts that the `phantom' density can be negative at places; so one may question the adding up of phantom masses (see, e.g., Ref. \cite{milgrom10}). The relevance of this effect is thus not clear, and is theory dependent.
\par
It is also important to realize that since the cosmological dark-matter anomaly is of order unity ($\z_c\approx 2\pi$), it is not necessarily directly connected with some low acceleration relative to $\az$. The anomaly could simply be some combination of the dimensionless parameters appearing in the eventual, more complete, MOND theory. For example, this is a demonstrated possibility for the cosmological dark-energy anomaly: In the relativistic, bimetric MOND (BIMOND) class of theories \cite{milgrom09}, a cosmological constant that is a dimensionless multiple of $\az^2/c^4$ appears, which is not a low-acceleration effect (where $\az$ appears in the theory in internal dynamics of galaxies). The dimensionless factor remains a free parameter of the theory, until we can found the theory on a deeper stratum.
In fact, in this BIMOND class of theories, there is enough choice of the dimensionless parameters to give a different value to Newton's $G$, as appears in local gravitational physics, and to $G_c$ that appears in cosmology, perhaps accounting for the cosmological dark-matter expansion anomaly. It remains to be seen, though, that this idea can be made to work in detail (for example, to circumvent the limit on $G_c/G$ in the very early Universe).


\begin{thebibliography}{0}

\bibitem{spergel15}%
  \textsc{D.\,E.~ Spergel},
   \jr{Science} \textbf{347}, 1100 (2015).
\bibitem{disney08}%
  \textsc{M.\,J.~Disney et al.}
  \jr{Nature} \textbf{455},  1082 (2008).

\bibitem{kroupa12}%
  \textsc{P.~ Kroupa},
   \jr{Pub. Astr. Soc. Austr.} \textbf{29 (4)}, 395 (2012).

\bibitem{riess16}%
  \textsc{A.\,G.~Riess et al.}
  \jr{} \textbf{},  arXiv:1604.01424 (2016).


\bibitem{milgrom83}%
  \textsc{M.~ Milgrom},
   \jr{ApJ} \textbf{270}, 365 (1983).


\bibitem{fm12}%
  \textsc{B.~ Famaey}, and
  \textsc{S.~ McGaugh},
  \jr{Liv. Rev. Rel.} \textbf{15}, 10 (2012).

\bibitem{milgrom14c}%
  \textsc{M.~ Milgrom},
  \jr{Scholarpedia} \textbf{9(6)}, 31410 (2014).

\bibitem{milgrom15a}%
  \textsc{M.~ Milgrom},
  \jr{Can. J. Phys.} \textbf{93(2)}, 107 (2015).
\bibitem{milgrom14}%
  \textsc{M.~ Milgrom},
  \jr{MNRAS} \textbf{437}, 2531 (2014).

\bibitem{milgrom15}%
  \textsc{M.~ Milgrom},
  \jr{Phys. Rev. D} \textbf{92}, 044014 (2015).
\bibitem{sn07}%
  \textsc{R.\,H.~Sanders}, and
  \textsc{E.~Noordermeer},
  \jr{MNRAS} \textbf{379}, 702 (2007).

\bibitem{milgrom09a}%
  \textsc{M.~ Milgrom},
   \jr{ApJ} \textbf{698}, 1630 (2009).

\bibitem{lueg15}%
  \textsc{F. L\"{u}ghausen et al.},
\jr{Can. J. Phys.} \textbf{93}, 232 (2015).

\bibitem{candlish15}%
  \textsc{G.\,N. Candlish et al.},
\jr{MNRAS} \textbf{446}, 1060 (2015).

\bibitem{milgrom83a}%
  \textsc{M.~ Milgrom},
   \jr{ApJ} \textbf{270}, 371 (1983).




\bibitem{sanders96}%
  \textsc{R.\,H.~Sanders},
  \jr{ApJ} \textbf{473}, 117 (1996).


\bibitem{mcgaugh12}%
  \textsc{S.\,S.~McGaugh},
  \jr{Astron. J.} \textbf{143}, 40 (2012).

\bibitem{papastergis16}%
  \textsc{E.~Papastergis et al.},
  arXiv:1602.09087  (2016).

\bibitem{denheijer15}%
  \textsc{M. den Heijer et al.},
\jr{AA} \textbf{581}, A98 (2015).


 \bibitem{milgrom13}%
  \textsc{M.~ Milgrom},
   \jr{Phys. Rev. Lett.} \textbf{111}, 041105 (2013).



\bibitem{sanders90}%
  \textsc{R.\,H.~Sanders},
  \jr{AA Rev.} \textbf{2}, 1 (1990).

\bibitem{scarpa06}%
  \textsc{R.~ Scarpa},
  \jr{AIP Conf. Proc.} \textbf{822}, 253, arXiv:astro-ph/0601478 (2006).

\bibitem{milgrom94}%
  \textsc{M.~ Milgrom},
   \jr{Ann. Phys.} \textbf{229}, 384 (1994).
\bibitem{mcgaugh14}%
  \textsc{S.\,S.~McGaugh},
  \jr{Galaxies} \textbf{2(4)}, 601 (2014).

\bibitem{bf16}%
  \textsc{L.~ Blanchet}, and
  \textsc{B.~ Famaey},
  \jr{}  arXiv:1602.00711 (2016).


\bibitem{oman16}%
  \textsc{K.\,A.~Oman et al.},
  \jr{} \textbf{}  arXiv:1601.01026 (2016).

\bibitem{kent87}%
  \textsc{S.\,M.~ Kent},
   \jr{Astron. J.} \textbf{93}, 816 (1987).

 \bibitem{begeman91}%
  \textsc{K.\,G.~Begeman et al.},
\jr{MNRAS} \textbf{249}, 523 (1991).

\bibitem{dm98}%
\textsc{W.\,J.\,G.~de Blok}, and
  \textsc{S.~ McGaugh}
   \jr{ApJ} \textbf{508}, 132 (1998).

 \bibitem{gentile11}%
  \textsc{G. ~Gentile et al.},
  \jr{AA} \textbf{527}, A76  (2011).



 \bibitem{haghi16}%
  \textsc{H. ~Haghi et al.},
  \jr{MNRAS} \textbf{ },   (2016).

    \bibitem{milgrom12a}%
  \textsc{M.~ Milgrom},
   \jr{Phys. Rev. Lett.} \textbf{109}, 131101 (2012).

\bibitem{serra10}%
  \textsc{A.\,L.~Serra et al.},
  \jr{AA} \textbf{524}, 13 (2010).

\bibitem{mm13}%
  \textsc{S.~ McGaugh}, and
  \textsc{M.~ Milgrom},
  \jr{ApJ} \textbf{775}, 139 (2013).

\bibitem{wu15}%
  \textsc{X.~ Wu}, and
  \textsc{P.~ Kroupa},
  \jr{MNRAS} \textbf{446}, 330 (2015).

   \bibitem{hees16}%
  \textsc{A. ~Hees et al.},
  \jr{MNRAS} \textbf{455 }, 449 (2016).

\bibitem{milgrom99}%
  \textsc{M.~ Milgrom},
  \jr{Phys. Lett. A} \textbf{253}, 273 (1999).

\bibitem{unruh75}%
  \textsc{W.\,G.~Unruh},
  \jr{Phys. Rev. D} \textbf{14}, 870 (1975).

\bibitem{sakharov68}%
  \textsc{A.\,D.~Sakharov},
  \jr{Sov. Phys. Dok.} \textbf{12}, 1040 (1968).

  \bibitem{jacobson95}%
  \textsc{T.~Jacobson},
  \jr{Phys. Rev. Lett.} \textbf{75}, 1260 (1995).

\bibitem{padmanabhan10}%
  \textsc{T.~Padmanabhan},
  \jr{Rep. Prog. Phys.} \textbf{73}, 046901 (2010).

\bibitem{verlinde11}%
  \textsc{E.~Verlinde},
  \jr{JHEP} \textbf{2011}, 29 (2011).


\bibitem{milgrom06a}%
  \textsc{M.~ Milgrom},
  \jr{Phys. Rev. D} \textbf{73}, 084005 (2006).

\bibitem{chiou16}%
  \textsc{D-W.~Chiou},
  \jr{} \textbf{}  arXiv:1605.06656 (2016).


\bibitem{pikhitsa10}%
  \textsc{P.\,V.~Pikhitsa},
  \jr{} \textbf{}  arXiv1010.0318 (2010).


 \bibitem{klinkhamer12}%
  \textsc{F.\,R.~ Klinkhamer},
   \jr{Mod. Phys. Lett. A} \textbf{27}, 1250056 (2012)


\bibitem{blanchet15}%
  \textsc{L.~ Blanchet}, and
  \textsc{L.~ Heisenberg},
  \jr{JCAP} \textbf{2015 (12)}, 26 (2015).

\bibitem{khoury16}%
  \textsc{J.~ Khoury},
    \jr{Phys. Rev. D} \textbf{93}, 103533 (2016).


  \bibitem{milgrom06}%
  \textsc{M.~ Milgrom},
  \jr{EAS Pub.} \textbf{20}, 217, astro-ph/0510117 (2006).

\bibitem{planck15}%
  \textsc{P.\,A.\,R.~Ade et al.},
  \jr{} \textbf{}  arXiv:1502.01589 (2015).

\bibitem{milgrom10}%
  \textsc{M.~ Milgrom},
  \jr{Phys. Rev. D} \textbf{82}, 043523 (2010).


\bibitem{nusser02}%
  \textsc{A. ~ Nusser},
   \jr{MNRAS} \textbf{331}, 909 (2002).

\bibitem{llinares11}%
  \textsc{C. ~ Llinares},
   \jr{PhD thesis, U. of Groningen} \textbf{}(2011).

\bibitem{candlish16}%
  \textsc{G.\,N.~ Candlish},
  \jr{MNRAS} \textbf{}, arXiv:1605.03192 (2016).


\bibitem{carroll04}%
  \textsc{S.\,M.~ Carroll}, and
  \textsc{E.\,A.~ Lim},
  \jr{Phys. Rev. D}  \textbf{70}, 123525 (2004).

 \bibitem{milgrom09}%
  \textsc{M.~ Milgrom},
  \jr{Phys. Rev. D} \textbf{80}, 123536 (2009).





\end{thebibliography}
\end{document}